\documentclass[RNAAS]{aastex62}

\newcommand{\quotes}[1]{``#1''}
\begin{document}

\title{WISE J064336.71-022315.4:\\
  A Thick Disk L8 Gaia DR2-Discovered Brown Dwarf at 13.9 Parsecs}

\correspondingauthor{Eric E. Mamajek}
\email{mamajek@jpl.nasa.gov}

\author[0000-0003-2008-1488]{E. E. Mamajek}
\affiliation{Jet Propulsion Laboratory, California Institute of Technology, M/S
  321-100, 4800 Oak Grove Drive, Pasadena, CA 91109, USA}
\affiliation{Department of Physics \& Astronomy, University of
  Rochester, Rochester, NY 14627, USA}

\author[0000-0001-7519-1700]{F. Marocco}
\affiliation{Jet Propulsion Laboratory, California Institute of Technology,
  M/S 321-100, 4800 Oak Grove Drive, Pasadena, CA 91109, USA}

\author[0000-0002-5376-3883]{J. M. Rees}
\affiliation{Center for Astrophysics and Space Sciences, University of California,
  San Diego, San Diego, CASS, M/C 0424 9500 Gilman Drive, La Jolla, CA 92093-0424, USA}

\author[0000-0002-4424-4766]{R. L. Smart}
\affiliation{Istituto Nazionale di Astrofisica, Osservatorio Astrofisico di Torino,
  Strada Osservatorio 20, I-10025 Pino Torinese, Italy} 

\author[0000-0003-3501-8967]{W. J. Cooper}
\affiliation{School of Physics, Astronomy and Mathematics, University of Hertfordshire,
  College Lane, Hatfield AL10 9AB, UK}
\affiliation{Istituto Nazionale di Astrofisica, Osservatorio Astrofisico di Torino,
  Strada Osservatorio 20, I-10025 Pino Torinese, Italy} 

\author[0000-0002-6523-9536]{A. J. Burgasser}
\affiliation{Center for Astrophysics and Space Sciences, University of California,
  San Diego, San Diego, CASS, M/C 0424 9500 Gilman Drive, La Jolla, CA 92093-0424, USA}


\section{}

While spectroscopically characterizing nearby ultracool dwarfs
discovered in the Gaia Second Data Release \citep[GDR2;][]{GaiaDR2}
with the TripleSpec spectrograph on the Palomar 200'' telescope, we
encountered a particularly cool, nearby, new member of the solar
neighborhood: Gaia DR2 3106548406384807680 = WISE J064336.71-022315.4
= 2MASS J06433670-0223130 (hereafter W0643).\\

{\it Astrometry:} \citet{GaiaDR2} reports an epoch 2015.5 ICRS
position for W0643 of $\alpha$, $\delta$ = $100$\fdg$90303036652$
($\pm$1.3 mas), $-2$\fdg$38793881136$ ($\pm$1.2 mas) and proper motion
$\mu_{\alpha}$, $\mu_{\delta}$ = $28.3\,\pm\,2.3$, $-221.2\,\pm\,2.5$
mas\,yr$^{-1}$.  The \citet{GaiaDR2} parallax $\varpi$ =
71.9\,$\pm$\,1.4 mas corresponds to a distance of 13.9\,$\pm$\,0.3 pc.
Combined, these yield a tangential velocity $V_{tan}$ =
14.7\,$\pm$\,0.4 km\,s$^{-1}$.\\

{\it Photometry:} W0643 has 2MASS photometry \citep{Skrutskie06} of
$J$ = 15.48\,$\pm$\,0.05, $H$ = 14.38\,$\pm$\,0.05, $K_s$ =
13.62\,$\pm$\,0.05 ($J-K_s$ = 1.86$\pm$0.07), and AllWISE photometry
\citep{2010AJ....140.1868W} of $W1$ = 12.85\,$\pm$\,0.03, $W2$ =
12.50\,$\pm$\,0.03, $W3$ = 11.54\,$\pm$\,0.26, and $W4$ $<$ 8.74.
Gaia DR2 reports $G$ = 20.680\,$\pm$\,0.014.  Pan-STARRS PS1
\citep{Chambers16} reports $i$ = 21.071\,$\pm$\,0.007, $z$ =
18.505\,$\pm$\,0.013, and $y$ = 17.474\,$\pm$\,0.013. Combined with
the Gaia DR2 parallax, we estimate absolute magnitudes $M_G$ =
19.96\,$\pm$\,0.04, and $M_{Ks}$ = 12.91\,$\pm$\,0.06. The colors and
magnitudes are typical for late-L dwarfs \citep{Wang18}.\\

{\it Spectroscopy:} W0643 was observed 12:47 UT 17 October 2018 with
TripleSpec on the Palomar 200" \citep{Herter08}, with a 1\arcsec
$\times$ 30\arcsec\, slit, and the spectrum covering 1.0-2.4 $\mu$m at
resolution R $\simeq$ 2600.  W0643 was observed at airmass 1.26 and
conditions were clear with $\sim$1\arcsec.5 seeing.  We obtained 8
frames of 240\,s each in an ABBA nodding pattern with the slit aligned
with the parallactic angle.  The A0V star HD~54601 was observed
afterwards to provide telluric absorption correction and flux
calibration \citep{Vacca03}.  Data were reduced with a modified
version of SpeXtool \citep{Cushing04}.\\


%

{\it Analysis:} In Fig. 1, we compare W0643's TripleSpec spectrum to
SpeX Prism Spectral Library standards using {\it SPLAT}
\citep{Burgasser14, Burgasser17}.  We classify W0643 as L8, with a
\quotes{plateau-shaped} $H$-band spectrum typical for field Ls
\citep{Allers13}.  We measured a heliocentric radial velocity of
142\,$\pm$\,12 km\,s$^{-1}$.  When combined with {\it Gaia}
astrometry, we determine a Galactic velocity (heliocentric; $U$
towards Galactic center) of $U, V, W$ = -109, -91, -12 ($\pm$10, 5, 3)
km\,s$^{-1}$.  We estimate that W0643 passed within $\sim$1.4 pc away
from the Sun $\sim$100,000 years ago.  Using the kinematic criteria of
\citet{Bensby03}, we estimate a 96\% probability that W0643 is a thick
disk star, which implies an age of 9-13 Gyr \citep{Haywood18}.\\

Using the VOSA SED analyzer
\citep{Bayo08}\footnote{\url{http://svo2.cab.inta-csic.es/theory/vosa/}}
with the Pan-STARRS/2MASS/WISE photometry, we find a best fit BT-Settl
CIFIST spectrum with T$_{\rm eff}$ = 1400\,K, log($g$) = 4.5, with
solar metallicity, with luminosity log(L/L$_{\odot}$) =
-4.61\,$\pm$\,0.02 dex.  Combining these values, the age constraint,
and the {\it Sonora 2018} evolutionary models \citep{Marley18}, we
predict W0643's mass to be $\sim$0.070 $M_{\odot}$ (i.e. a brown
dwarf).\\

\begin{figure}[htbp!]
\begin{center}
\includegraphics[scale=0.9,angle=0]{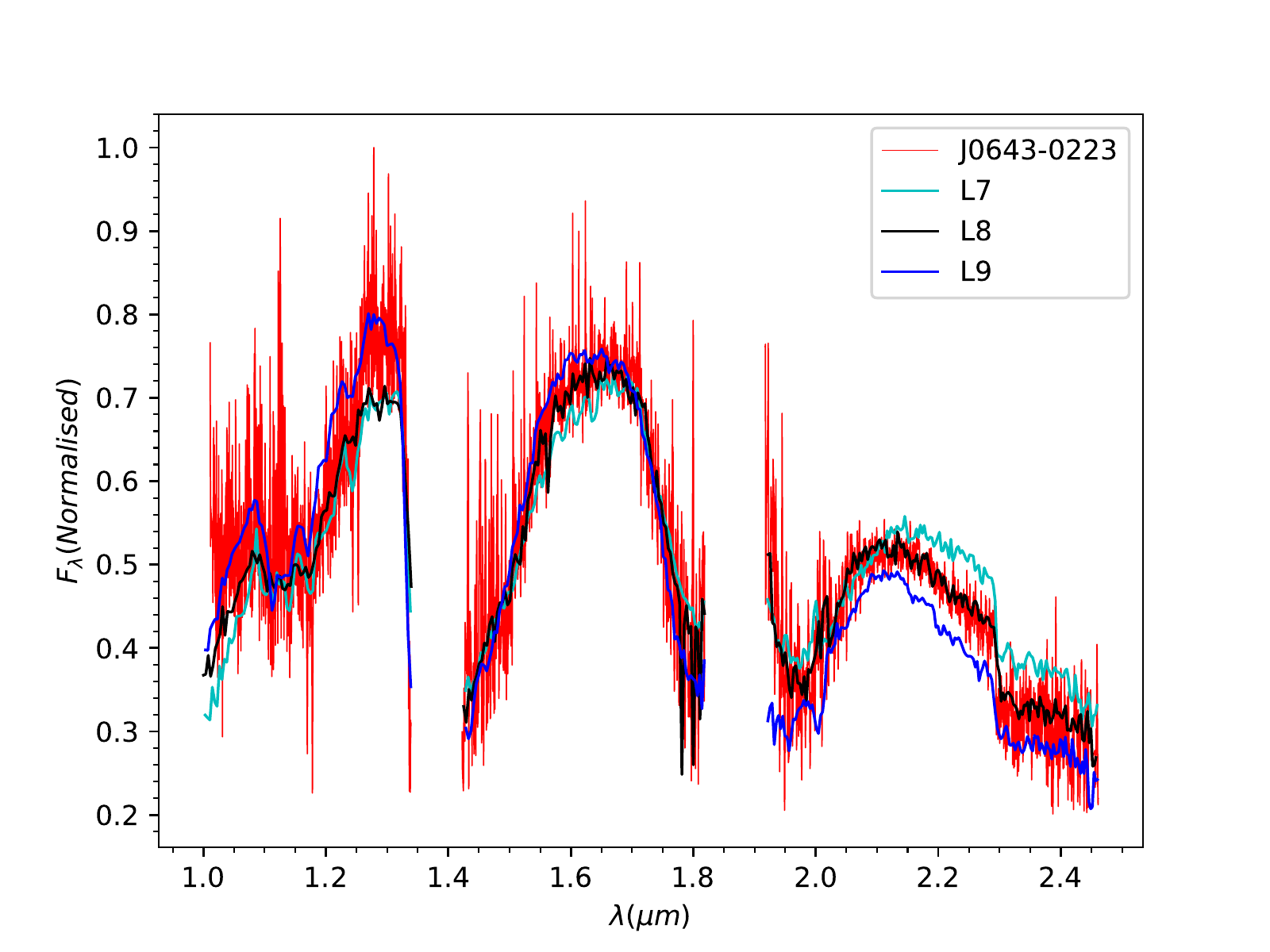}
\caption{TripleSpec $JHK$-band spectrum of WISE J064336.71-022315.4
  (red line) compared to those of the L dwarf standards 2MASS
  J0103+1935 \citep[L7, in teal;][]{Cruz04}, 2MASS J1632+1904
  \citep[L8, in black;][]{Burgasser07} and DENIS J0255-4700 \citep[L9,
    in blue;][]{Burgasser06}.}
\end{center}
\end{figure}

\acknowledgments

Part of this research was carried out at the Jet Propulsion
Laboratory, California Institute of Technology, under a contract with
NASA.
This work has made use of data from the European Space Agency (ESA) 
Gaia mission (\url{https://www.cosmos.esa.int/gaia}), 
processed by the Gaia Data Processing and Analysis Consortium 
(DPAC, \url{https://www.cosmos.esa.int/web/gaia/dpac/consortium}).
This research has benefitted from the SpeX Prism Spectral Libraries,
maintained by Adam Burgasser at
\url{http://pono.ucsd.edu/~adam/browndwarfs/spexprism}.\\

\facilities{Gaia, Hale, WISE, 2MASS, PS1, IRTF}

\end{document}